\documentclass[preprint,aps,showpacs]{revtex4}
\usepackage{longtable} 
\usepackage{graphics}

\begin{document}

%%-move to normal A4-%%

\title{Charge induced effects on structures and properties of 
silane and disilane derivatives} 

\author{ D. Balamurugan}  
 \altaffiliation[Present address: ]{Department of Chemistry, 
Tulane University, New Orleans, LA 70118}   
 \affiliation{Department of Physics, Indian Institute of Technology, 
Kanpur 208016, India}
\author{R. Prasad}  
\affiliation{Department of Physics, Indian Institute of Technology, 
Kanpur 208016, India}

\begin{abstract}
Using \emph{ab-initio} electronic structure methods we
have investigated  the ground state structures and properties of
neutral and charged SiH$_{m}$(m=1-4) and Si$_{2}$H$_{n}$(n=1-6) 
clusters which are produced in the plasma enhanced  
chemical vapor deposition process used in the preparation of hydrogenated
amorphous silicon({\it{a}}-Si:H).  Our results show that 
charging a neutral cluster distorts it and the distortion mainly occurs 
through the orientation of Si-H bond. We attribute  structural changes 
in the charged clusters to  electrostatic repulsion between the bonded  
and non-bonded electrons.  We find that in addition 
to the usual Si-H bond, hydrogen atoms    
form Si-H-H and Si-H-Si bonds in some clusters.  The vibrations  of
 Si-H, Si-Si, Si-H-Si bond stretching modes show that the 
frequencies  are  shifted significantly    
upon charging. The  frequency shifts in the charged 
clusters are  consistent with their bond length variations.  We 
discuss the fragmentation pathways of 
silane into binary products and the role of fragmented silane 
radicals in the cluster formation and {\it{a}}-Si:H film 
deposition process. 

\end{abstract}
\pacs{73.22.-f, 31.15.Ar, 36.40.Wa, 36.40.Mr}
\maketitle

\section{Introduction}

Silane(SiH$_{4}$)  is used 
in the plasma enhanced chemical vapor deposition(PECVD) process to produce
hydrogenated amorphous silicon({{\it{a}}-Si:H)  which has potential
applications in solar cells and flat panel devices. Before {\it{a}}-Si:H film 
deposition takes place, the gaseous silane molecules are decomposed
either by  using 
electric field\cite{KesselsJAP01v89} or by thermal 
energy\cite{DuanAPL01v78}. The decomposed species live for a short time 
in  plasma environment and contribute to {\it{a}}-Si:H film
growth. In the plasma, the 
decomposed species exist also in charged states.
Experimental 
studies\cite{DuanAPL01v78, LerouxJAP00v88, GallagherJAP05v98, 
GallagherJAP04v96, GallagherJAP02v91, GallagherJAP00v87, 
GallagherJAP92v71, GallagherJAP92v71b, GallagherJAP90v68, GallagherPRA90v42} have shown the presence of    
various neutral and charged species in the plasma such as  
Si, SiH$_{3}$ and Si$_{2}$H$_{6}$ and 
SiH$_{3}$$^{+}$,  Si$_{2}$H$_{5}$$^{+}$ and Si$_{x}$H$^{+}$ (x=4-10). Also 
the experimental 
investigations\cite{KeudellPRB99v59, KesselsJAP01v89, GallagherJAP05v98,
GallagherJAP02v91, GallagherJAP97v82, GallagherAPL96v68} show that 
the decomposed species have a great influence  on the properties
of  {\it{a}}-Si:H films. Therefore, investigating the structure and 
properties of  neutral and charged silane and disilane derivatives 
can help in gaining a better understanding of {\it{a}}-Si:H film growth in 
the PECVD process. With this  
motivation  we have carried out \emph{ab-initio} electronic structure 
calculations to study  (1) the ground state structure,  (2) vibrational 
frequencies, (3) fragmentation and (4) clustering process 
of  neutral and charged SiH$_{m}$(m=1-4) and Si$_{2}$H$_{n}$(n=1-6) 
clusters. 

In the recent past, several electronic structure calculations 
have been presented on neutral and charged 
silane and disilane derivatives  at various levels of 
accuracy\cite{KalemosJCP02v116, Shen90v93, AllenCP86v108,Grev92v97, GupteIJMP98v12, PakJCP03v118}.
 The ground and excited states of
SiH have been investigated using multireference 
configuration interaction approach\cite{KalemosJCP02v116}. 
Using configuration interaction and coupled cluster method 
with single and double excitations,   
a detailed vibrational analysis on  
SiH$_{3}$$^{-}$ has been performed\cite{Shen90v93}. Coupled cluster 
calculations have been carried out to study the  
ground state structures and vibrational properties of SiH$_{n}$ (n = 2-4) 
clusters\cite{AllenCP86v108}.  Grev and Schaefer\cite{Grev92v97}  applied
coupled cluster method to identify the local minima of    
Si$_{2}$H$_{2}$.  Gupte 
and Prasad\cite{GupteIJMP98v12} reported ground state structures and 
vibrational frequencies of Si$_{n}$H$_{m}$ (n=1,2 and m = 1-6) clusters 
using non-orthogonal tight binding molecular dynamics. 
Recently,  the potential energy surface of Si$_{2}$H$^{-}$ has been 
explored\cite{PakJCP03v118} using coupled cluster and configuration 
interaction  methods. However, \emph{ab-initio} calculations on 
neutral and charged 
silane and disilane derivatives using the same methodology are  
still lacking. We have performed density functional theory 
calculations  to understand the charge induced effects on the structure and 
properties of silane and disilane derivatives.    

The fragmentation and the clustering reactions of silane derivatives  
in the PECVD process are not yet well 
understood. Phenomenological model calculations have been carried out to 
understand the formation of various clusters during silane discharge 
and their role on film deposition 
process\cite{GallagherPRE00v62,GallagherJAP00v87}. Using 
semi-empirical method
the collisions between Si$^{+}$ and H$_{2}$ over a range of collision 
energies have 
been studied\cite{ChaabaneJNCS02v299}. Sosa and 
Lee\cite{SosaJCP93v98}  performed density functional theory calculations 
to study  the insertion reactions of 
SiH$_{2}$, SiHF and SiF$_{2}$ into H$_{2}$ and 
found that fluorine substitution increases the 
reaction barrier heights.  In a series of theoretical 
investigations,  Raghavachari\cite{RaghavachariJCP90v92, RaghavachariJCP91v95, 
RaghavachariJCP92v96, RaghavachariJCP88v88}  showed that the 
sequential clustering 
reactions of silane with Si$^{+}$, SiH$^{+}$, SiH$_{2}$$^{+}$ 
and SiH$_{3}$$^{+}$ 
 end with different products such as Si$_{2}$H$_{4}$$^{+}$, 
Si$_{2}$H$_{5}$$^{+}$,  
Si$_{3}$H$_{3}$$^{+}$, Si$_{3}$H$_{5}$$^{+}$, 
Si$_{4}$H$_{7}$$^{+}$,  Si$_{5}$H$_{10}$$^{+}$ and  
Si$_{5}$H$_{11}$$^{+}$.  We have also investigated the fragmentation of 
silane and the role of the fragmented radicals in the beginning steps of 
the clustering process.

Our \emph{ab-initio} calculations on neutral and 
charged SiH$_{m}$(m=1-4) and Si$_{2}$H$_{n}$(n=1-6) clusters show
several interesting findings about their ground state 
structures and properties.  We find that 
in addition to the  usual Si-H bond, hydrogen atoms in some 
clusters form Si-H-H  and Si-H-Si  bonds. Charging the cluster affects the 
Si-H bond orientations and Si-H, Si-H-Si and Si-Si bond 
lengths. We explain the bond orientations and bond length variations 
of the charged clusters using molecular orbital, charge density and 
population analysis

We have done vibrational analysis of the neutral and charged clusters
and find that the frequencies of stretching modes of  
Si-H, Si-Si and H-Si-H bonds  shift considerably upon 
charging. These frequency shifts   
are consistent with the bond length 
variations of the charged clusters.  The fragmentation studies 
on neutral and charged silane show that 
the major radicals formed are SiH$_{2}$, SiH$_{2}$$^{+}$, SiH$_{3}$$^{+}$ 
and SiH$_{3}$$^{-}$. These radicals undergo further clustering 
processes which lead to disilane derivatives.  We also discuss the possible 
role of these  radicals  in {\it{a}}-Si:H film deposition process.

The plan of the paper is as follows. In section \textbf{II}
we give computational details of the present work. In
section \textbf{III} we present and discuss in detail 
the ground state geometries of neutral and charged  
silane and disilane derivatives.  Vibrational analysis of these clusters is  
discussed in section \textbf{IV}.  In section \textbf{V} we discuss 
the fragmentation of silane and  the formation of  
bigger 
hydrogenated silicon clusters.  Finally, in section \textbf{VI} we 
give our conclusions. 

\section{Computational details}

We have used the Car-Parrinello molecular 
dynamics(CPMD)\cite{CarPRL85v55,ParrinelloSSC97v102} 
and 
Gaussian98\cite{Gaussian98} package 
for our study. Using the 
CPMD method\cite{foot1, BalaPRB01v64} with simulated 
annealing optimization we have obtained the ground state structures 
of the neutral clusters.  The pseudopotentials for silicon
and hydrogen
have been generated using the  Bachelet, Hamann
and Schl$\ddot{u}$ter technique\cite{BacheletPRB82v26}. The local density
approximation(LDA) of the density functional theory
has been used with the Ceperley-Alder\cite{CeperleyPRL80v45} 
exchange-correlation
energy  functional parameterized by Perdew and 
Zunger\cite{PerdewPRB81v23}.
The wave functions were expanded in a plane wave basis with  12 Rydberg
energy cut-off and \textbf{k}=0 point was
used for Brillouin zone sampling. During  simulation volume of the
system was kept constant and to avoid  interaction between the
clusters a big fcc supercell with side length of 35 a.u.
was used.  To perform simulated annealing, the system was
taken to high temperatures (1200K in the steps of 300K), equilibrated
for a long time (about 16,000 steps)  and
then slowly cooled down (in the steps of 50K)
to 300K. Below this temperature the steepest descent optimization
was found to be  more efficient to obtain the
ground state geometry. The
desired temperature was  achieved by rescaling
 atomic velocities and the  atoms were  moved according to the velocity
Verlet algorithm\cite{Haile} with a time step of 5 a.u. The
fictitious mass of the electron
was  taken  to be  200 a.u.
Our SiH$_{n}$(n = 2-4) and Si$_{2}$H$_{2}$ 
ground state structures obtained using the CPMD method are in good 
agreement with those found  earlier\cite{AllenCP86v108, Grev92v97}
using \emph{ab-initio} quantum chemical calculations.

For obtaining the ground state structures of the 
charged clusters\cite{foot2} we have
used Gaussian98 package.  We have taken the structures of 
neutral clusters obtained from the CPMD calculations and re-optimized 
the neutral and charged clusters using Gaussian98 package by   
Berny's optimization technique\cite{SchlegelJCC82v3}. 
The geometry 
optimization has been done with 
Becke's 3 parameter hybrid
functional(B3LYP) for  
exchange-correlation functional and employing 
6-311g** basis set\cite{Gaussian98, Foresman}. The  
vibrational frequencies  of the neutral and 
charged clusters are calculated  with the same  functional 
and  basis set.  We see from Table I that the ionization potentials (IP)
of the clusters obtained from Gaussian calculations    
are in good agreement with the experimental 
values\cite{BerkowitzJCP87v86, ItohJCP86v85,
RobertsonJAP86v59}. The IP values of 
Si and Si$_{2}$ have been calculated with various 
multiplicities to take account of the orbital degeneracy. We find that the ground 
state of Si, Si$^{+}$, Si$_{2}$ and Si$_{2}^{+}$  have 
multiplicities 3, 2, 3 and 4 respectively. The IP values of
Si and Si$_{2}$ reported in Table I have been obtained using
the  ground state multiplicities. 
Our  calculated umbrella mode 
frequency 894 cm$^{-1}$ 
of SiH$_{3}$$^{-}$  is in excellent agreement with the 
experimental value\cite{NimlosJACS86v108} of 880 cm$^{-1}$.  

\section{Ground state structures}
\subsection{ Charge induced structural modifications}

In this section we discuss the charge induced effects on the 
structures of SiH$_{m}$(m=1-4) and Si$_{2}$H$_{n}$(n=1-6) 
clusters. In Fig. 1 and 2 we have  shown the ground state structures 
of neutral and charged silane and disilane derivatives obtained
using Gaussian98 package employing B3LYP/6-311G**.  From Fig. 1  
we see that the structure of  SiH$_{2}$ is  asymmetrical top whose 
bond angle opens up when 
positively charged and does not change much  when 
negatively charged.   The  
structure of SiH$_{3}$ is pyramidal which becomes planar  when 
positively charged and Si-H bonds further bend when negatively 
charged.  

The ground state structure of SiH$_{4}$  has 
tetrahedral symmetry. This symmetry is destroyed upon
charging\cite{foot3, BalaPRA03v69,ItohJCP86v85}and we see in Fig. 1 
that SiH$_{4}$$^{+}$ and SiH$_{4}$$^{-}$ have very different structures.   
This symmetry breaking has been attributed to  the Jahn-Teller 
effect\cite{BalaPRA03v69,BersukerCM01v101, solidv20, Englman}.The 
structure of SiH$_{4}$$^{+}$ obtained by our calculation is  
in agreement with the structure obtained by earlier 
calculations\cite{FreyJCP88v89,ProftCPL96v262,KudoCP88v122, 
CaballolCPL86v130,PorterJCP01v114}.  Our ground state of 
 SiH$_{5}$$^{+}$ is in good agreement with the 
result obtained by Boo and Lee\cite{BooJCP95v103}.  The structure  
of SiH$_{5}$$^{+}$ is more  
like a complex of SiH$_{3}$$^{+}$ and H$_{2}$  and that of   
SiH$_{5}$$^{-}$ is more like a complex of SiH$_{4}$$^{-}$ and H.  

Interestingly,  two H atoms in SiH$_{4}$$^{+}$ and SiH$_{5}$$^{+}$ 
are over-coordinated and form Si-H-H bond.  This Si-H-H 
bond is  formed between three atoms by two electrons and is an 
example of multi-center bond\cite{MarxSci99v284}. A similar type of 
three center-two electron  bond, C-H-H bond, has been 
observed\cite{MarxSci99v284, MarxAng97v36}  in   
CH$_{5}$$^{+}$. It was predicted that due to 
the three center - two electron bond,  
SiH$_{5}$$^{+}$ and CH$_{5}$$^{+}$
have fluxional behavior\cite{MarxSci99v284, MarxAng97v36, BooJCP95v103}, 
that is, these molecules do not have definite structure and bonding 
arrangement which makes it difficult to characterize them using 
spectroscopic techniques. These results show that charging clusters 
not only changes  the structure but also can give rise to the formation 
of unusual bonds.   

The over-coordination of H atom has also been seen in neutral clusters 
such as  Si$_{2}$H, Si$_{2}$H$_{2}$ and Si$_{2}$H$_{3}$.  We see 
from Fig. 2  that  hydrogen atom is over-coordinated in  
these clusters by connecting both silicon atoms and forms a Si-H-Si 
bridge bond.  Note that Si-H-Si bridge bond is a multi-centered bond 
and is different from Si-H bond, which is a two-centered 
bond\cite{BalaPRB01v64}. The Si-H-Si 
bridge bonded frame of the clusters does not change when the 
clusters are charged.  In the case of 
negatively charged  Si$_{2}$H$_{2}$, hydrogen atoms are slightly  
pushed away from each other without changing each Si-H-Si frame.  

Now we discuss the charge induced effects on the structure of 
disilane derivatives.  From Fig. 2  we see 
that Si-H bonds of Si$_{2}$H$_{3}$ are tilted away from 
Si-Si bond when positively charged and toward Si-Si bond when negatively  
charged.  Si$_{2}$H$_{4}$ is nearly  planar and 
results  in a perfectly planar structure when positively 
charged.  On the other hand,  Si-H bonds  of Si$_{2}$H$_{4}$ 
tilt significantly out 
of  plane when negatively charged.  Si$_{2}$H$_{5}$ 
consist of SiH$_{3}$ and SiH$_{2}$ units  connected by 
Si-Si bond.  We see from the figure that  H atoms of 
SiH$_{2}$ unit in Si$_{2}$H$_{5}$$^{+}$ move slightly upward while 
those in Si$_{2}$H$_{5}$$^{-}$  are 
further pushed down.  From the figure we see that Si$_{2}$H$_{6}$ 
consist of two 
SiH$_{3}$ units  connected by Si-Si bond.  Upon charging there is no 
significant change 
in the overall structure,  but 
each SiH$_{3}$ units of Si$_{2}$H$_{6}$$^{+}$ and Si$_{2}$H$_{6}$$^{-}$ 
distort as those of 
SiH$_{3}$$^{+}$ and SiH$_{3}$$^{-}$ clusters. The ground states of 
neutral and 
charged silane and disilane derivatives show that the structural 
modifications occur mainly  through the bending of Si-H bonds.

The charge induced structural distortions of silane and disilane derivatives 
can be understood from the nature of valance electron density
distribution. First, we discuss the nature of valance electron distribution 
and its influence on  charged  
SiH$_{2}$ structure.  Then, using the same arguments we explain 
the structural modifications of other charged clusters. The valance 
electron  density distribution of neutral and charged SiH$_{2}$ are shown in 
Fig. 3. From the 
figure we see that the electron density of SiH$_{2}$ is highly localized 
near H atoms and also on top of silicon atom. We call the electron 
density between Si and H as bonded while on top of Si as 
non-bonded.  This non-bonded
electron distribution is  in
 some sense, similar to the Lewis dot\cite{Levine} of unpaired
electrons.  The density of bonded electrons is  
highly localized near H atoms due to the 
fact that  hydrogen is more 
electronegative\cite{RoblesJACS84v106} and withdraws electron from 
silicon atom.  We 
see from the figure that the 
non-bonded electron density 
of SiH$_{2}$, on top Si atom,  decreases when positively charged  and  
increases when negatively charged. Thus in Fig. 1 for SiH$_{2}$ the non-bonded
electron density is schematically represented by a half filled circle on Si.
For SiH$_{2}$$^{+}$ 
it is shown by an open circle indicating the decrease in the 
non-bonded electron density 
while for SiH$_{2}$$^{-}$, it is shown by a filled circle indicating 
the increase
in the non-bonded electron density.   

Now we discuss how the structural modifications of charged SiH$_{2}$ can be 
understood in terms of the bonded and non-bonded electron density. Since 
the ground state of SiH$_{2}$ is not a linear geometry, the non-bonded electron
density would exert a downward force on the bonded electron density. Since the
non-bonded electron density of SiH$_{2}$$^{+}$ is 
lower than that of neutral SiH$_{2}$, its downward push 
on H atoms of SiH$_{2}$$^{+}$ 
is also lower than that of neutral SiH$_{2}$ which results in 
the opening of H-Si-H bond angle.  From this argument  one
would expect that SiH$_{2}$$^{++}$ should have  further opening in the 
bond angle. Indeed, we find that the optimization on 
SiH$_{2}$$^{++}$ results in an almost linear geometry with a larger 
bond angle. In contrast, in SiH$_{2}$$^{-}$ there is increment in the non-bonded electron 
density on top of silicon atom. Thus one would expect 
that the bond angle should 
get smaller than that of the neutral cluster due to the excess push  on 
H atoms.  But the reduction of H-Si-H  angle is opposed  by the increased 
H-H repulsion.  The H-H repulsion
in SiH$_{2}$$^{-}$ increases because charges on H increase when SiH$_{2}$ 
is negatively charged\cite{foot4}.

 The structural modifications of other charged clusters can be
explained using  
the same arguments given in the previous paragraph.
To understand the nature of valance 
electron distribution, we have done charge density analysis 
of all these clusters.  We find that  
the non-bonded electron density distribution  
decreases when positively charged 
and increases when negatively charged. From the charge density 
analysis we have also located the non-bonded electron
density in these clusters.  As in the case of neutral and charged
SiH$_{2}$ clusters, the non-bonded 
electron density is schematically represented by 
half, open and filled circles
for the neutral, positive  and negative clusters 
in Fig. 1 and 2.  From 
this  schematic representation it is easy to understand the structural 
modifications in the charged clusters. Hydrogen atoms bonded to Si 
are attracted to the nearby open circle while they are 
 repelled away from the nearby 
filled circle. The attraction of H to an open circle and the repulsion of H to 
a filled circle are due to the electrostatic interaction as explained in 
the case of charged SiH$_{2}$ clusters.

\subsection{ Charge induced bond length variations}

We find that charging the clusters not only changes Si-H bond 
orientations but  
also Si-H, Si-H-Si and Si-Si bond lengths. In 
Fig. 4(a), 4(b) and 4(c),  
 we have shown Si-H, Si-H-Si and Si-Si  bond lengths of the neutral and 
charged clusters as a function of cluster size.  We see from the figure   
that Si-H bond length of a positively charged cluster is shorter 
than that of the corresponding
neutral cluster while it is longer for  the negatively charged 
cluster. On the other hand,   
Si-H-Si and Si-Si bond lengths of both  
positive as well as  negative charged cluster  are longer than that of 
the neutral cluster.  However, Si$_{2}$H$_{3}$$^{+}$ 
is an exception  where  Si-Si and 
Si-H-Si bond lengths are shorter  than those of the neutral cluster.  
>From Fig. 4(c) we see  that Si-Si 
bond length of 
a neutral Si$_{2}$H$_{n}$ cluster increases as a function of H and reaches 
maximum for Si$_{2}$H$_{6}$, but the bond length of  Si$_{2}$H$_{4}$ 
is shorter than that of its neighboring  
Si$_{2}$H$_{3}$ and Si$_{2}$H$_{5}$ clusters. 

To understand the Si-Si and Si-H bond length variations of charged 
silane and disilane 
derivatives we first consider the bond length variations of  
Si$_{2}$ and SiH dimers.  Si$_{2}$ has bond length of 2.144$\AA$, the 
bond expands in both positive as well as negative 
charged state  to 2.308 and 2.453$\AA$ respectively. 
On the other hand,
depending on the charged state the bond length of  SiH dimer  
decreases or increases.  SiH dimer has bond length of  
1.536$\AA$, which shortens to 1.518$\AA$ when positively charged  
 and expands to 1.566$\AA$ when negatively charged.  The different 
trends seen in the bond
lengths of  charged
Si$_{2}$ and SiH dimers can be understood
with the results obtained from population analysis. The population 
analysis on charged dimers are done by
keeping the bond distance same as those of neutral dimers.  
 The population analysis on Si$_{2}$
shows that each atom of the neutral dimer has zero charge, whereas
each atom of Si$_{2}$$^{+}$ has +0.5e and that of Si$_{2}$$^{-}$
has -0.5e. The atoms of charged Si$_{2}$ are
equally charged and repel each other which results in
bond length expansion in both positive and negative charged state.  The
population analysis on  neutral SiH dimer shows that a net charge of
+0.35e  on Si and -0.35e on H atom. The charge transfer from silicon
to hydrogen is the indication of  polar covalent\cite{Beiser} bond which 
is consistent considering the fact that electronegativity of silicon
is smaller than hydrogen\cite{RoblesJACS84v106}. The population
analysis on SiH$^{+}$ shows that the silicon atom in the dimer holds a
charge of +1.36e and hydrogen -0.36e. This means that the lost electron
is primarily
from Si atom in the SiH dimer. Hence the positively charged silicon pulls
the  H atom  closer and reduces the bond length.  In the
negative charged state, Si atom carries a charge of -0.68e and H atom,
 a charge of -0.32e and this shows that the extra electron is
accumulated primarily on silicon atom. This extra electron density on Si
of SiH$^{-}$ repels the H atom and results in the bond length expansion.
Our population analysis on neutral and charged SiH dimer shows that
Si atom holds the excess charge in SiH$^{+}$ and SiH$^{-}$. This can be 
understood by considering the energies required  
to add or remove an electron from Si and H atoms. Si atom has 
smaller ionization potential ( IP ) than H atom  
while it has larger electron affinity ( EA )  than H 
atom\cite{crchandbook}. Therefore adding or removing
an electron from Si is easier with respect to H atom. 

The general trends of Si-Si 
and Si-H bond length variations of 
charged clusters are consistent with the bond length variations 
of Si$_{2}$ and SiH dimers when charged. As mentioned earlier, 
Si-H-Si bond expands under both the charged state. In Si-H-Si unit, 
two silicon atoms are also bonded and they expand irrespective 
of the charged state. As a result of Si-Si expansion, Si-H-Si 
bridge bond expands under positive as well as negative charged state.  

The shorter Si-Si bond observed in Si$_{2}$H$_{4}$ is 
due to the formation 
of a $\pi$ bond between silicon atoms. The $\pi$ bond has a shorter 
bond length when compared to a $\sigma$ bond\cite{pibond}.  The highest 
occupied molecular orbital(HOMO) of Si$_{2}$H$_{4}$, Si$_{2}$H$_{3}$ and 
 Si$_{2}$H$_{3}$$^{+}$  are  
shown in Fig. 5. From the 
figure we see that the HOMO of Si$_{2}$H$_{4}$ is spread above 
and below the line connecting the
two Si atoms. This indicates that the HOMO of Si$_{2}$H$_{4}$ is formed due to 
the side overlap between valance {\it{p}} orbitals of silicon atoms which 
is a characteristic of $\pi$ bonding\cite{pibond}. Si$_{2}$H$_{3}$$^{+}$ has 
shorter Si-Si and Si-H-Si bonds  compared to the corresponding 
neutral cluster. From Fig. 5 we see that the HOMO of Si$_{2}$H$_{3}$$^{+}$ 
is slightly spread above and below the Si-Si bond while such a spread is
absent in neutral Si$_{2}$H$_{3}$. This means that HOMO of 
Si$_{2}$H$_{3}$$^{+}$ 
has gained slightly $\pi$ character compared to that of 
neutral Si$_{2}$H$_{3}$ and   
hence Si-Si bond is shorter than that of the neutral cluster. Due to the
reduction in Si-Si bond length, Si-H-Si bond length of 
Si$_{2}$H$_{3}$$^{+}$ is also reduced from that of the neutral cluster. 

\section{ Vibrational analysis} 

In this section we discuss the charge induced effects on  
vibrational frequencies of  the clusters.  Also we identify  
the modes corresponding to internal rotation of subunits in the 
clusters. The vibrational frequencies and 
force constants of silane and disilane derivatives are given in 
Tables II and III.  Our
calculated vibrational frequencies 922, 980, 2235 and 2244cm$^{-1}$ for 
SiH$_{4}$ are in good agreement with those of  experimentally obtained
values\cite{CoatsJMol94v320} of 914, 953, 2189 and
2267cm$^{-1}$.

We notice from the tables that  
 SiH$_{4}$$^{+}$, SiH$_{5}$$^{+}$, Si$_{2}$H$_{3}$$^{+}$, 
Si$_{2}$H$_{4}$$^{-}$, neutral and charged  
clusters of Si$_{2}$H$_{5}$ and Si$_{2}$H$_{6}$  
have one mode  with very low frequency and   
nearly zero force constant. A small frequency
and force constant for a mode imply  that during the 
vibrational motion  the atoms of the cluster move in a nearly constant
potential energy surface.  This may be an indication  
that the  mode corresponds to either transition state or internal 
rotation\cite{BalaPRA03v69, AyalaJCP98v108, QuadeJMS98v188, DuanCP02v280} 
of the cluster. We looked into the vibrational 
motion corresponding to low frequency modes of the clusters. 
We find that the low frequency  
modes  are associated with the  internal rotation of subunits about 
some symmetry axis of the clusters.

One vibrational frequency of SiH$_{3}$ corresponds to the 
umbrella mode. The umbrella mode frequency of SiH$_{3}$ is 760cm$^{-1}$ 
which increases  
for both  SiH$_{3}$$^{+}$ and SiH$_{3}$$^{-}$ to    
839 and 894 cm$^{-1}$ respectively. The inversion barriers for 
SiH$_{3}$ and SiH$_{3}$$^{-}$ are 0.22 and 1.24 eV respectively.
 Because of the increased barrier height, the 
frequency of SiH$_{3}$$^{-}$ is larger than that 
of SiH$_{3}$.  Although SiH$_{3}$$^{+}$ is planar without inversion 
barrier,  the frequency and force constant of the umbrella mode are 
larger than those of  the neutral and negative clusters. This is due to  
the fact that the  potential energy surface of 
SiH$_{3}$$^{+}$ about the center of inversion has different shape from  
those of SiH$_{3}$ and SiH$_{3}$$^{-}$. At the inversion 
center, SiH$_{3}$ and SiH$_{3}$$^{-}$ have  barriers while  
SiH$_{3}$$^{+}$ has potential energy minimum as shown in Fig. 6.

We find that charging the cluster shifts the stretching frequency of  
Si-H, Si-Si and H-Si-H bonds.  From Table II and III we see that the 
trend of frequency variations in Si-H, Si-Si and Si-H-Si stretch modes
of charged clusters is opposite to the trend of their bond length 
variations which is shown in Fig. 4.  For example,  Si-H stretch  
frequency of SiH dimer is 2013 cm$^{-1}$ which increases to 
2137 cm$^{-1}$ for SiH$^{+}$  and  decreases to 1827 cm$^{-1}$ for 
SiH$^{-}$.  From Fig. 4 we see that the bond length  is shorter for 
SiH$^{+}$ while it is longer for SiH$^{-}$ from that of 
the neutral SiH dimer. Such a relationship between Si-H bond length variation
and frequency variation seen in SiH dimer holds true for 
Si-H, Si-Si and Si-H-Si bonds of all the clusters. In general, longer  
bond length reduces the frequency and vice versa.  This relationship 
is reasonable since expansion of the bond means that the bond gets 
weaker  and hence the corresponding stretch mode has lower frequency. On 
the other hand  shortening of the bond length means that the bond gets 
stronger  and hence the stretch mode has larger frequency.

\section{ Fragmentation of silane and the formation of bigger clusters}

\subsection{Fragmentation of neutral and charged silane}

For the fragmentation studies, we have considered   various possible 
fragmentation pathways  of  silane into binary products.  Also 
the charged clusters are included in our investigation as they exist  
in PECVD plasma.  The fragmentation energy of a cluster 
has been calculated by taking the    
difference between total energy of reactant with those of  
product clusters. Among the possible fragmentation channels 
of neutral and charged silane the  low energy channels are   

\begin{equation}
SiH_{4} \rightarrow SiH_{2} + H_{2} ( +2.57 eV ) 
\end{equation}
\begin{equation}
SiH_{4}^{+} \rightarrow SiH_{2}^{+} + H_{2} ( +0.60 eV )  
\end{equation}
\begin{equation}
SiH_{4}^{-} \rightarrow SiH_{3}^{-} + H ( +1.33 eV )  
\end{equation}

where the number in the parenthesis is the fragmentation energy. 

As we see from the above reactions, production of hydrogen in molecular 
form is favored in SiH$_{4}$ and SiH$_{4}$$^{+}$ while  
atomic hydrogen  is favored in 
SiH$_{4}$$^{-}$.  Our low energy fragmentation channel of SiH$_{4}$ agrees 
with the experimental study\cite{GallagherJAP92v71}.  
There is another process where  hydrogen 
atom is attached with SiH$_{4}$$^{+}$ and forms SiH$_{5}$$^{+}$, which   
undergoes fragmentation as given below.  

\begin{equation}
SiH_{4}^{+} + H \rightarrow SiH_{5}^{+} \rightarrow 
SiH_{3}^{+} + H_{2} ( +0.64 eV )  
\end{equation}

This process is quite possible since a large amount of hydrogen is 
available in the PECVD plasma and  we see from the fragmentation energies 
of SiH$_{4}$$^{+}$ 
and SiH$_{5}$$^{+}$ that the latter is more stable to the fragmentation of   
H$_{2}$.  We find that the  energy required to break H$_{2}$ from 
SiH$_{5}$$^{-}$ is negative and is about -0.24 eV. This 
indicates that SiH$_{5}$$^{-}$ is highly unstable.
 Therefore, SiH$_{5}$$^{-}$ is not considered  in our 
fragmentation study. 

It has been observed that deposition of SiH$_{3}$ on the film improves 
the material quality whereas deposition of SiH and SiH$_{2}$  degrades  
the material quality\cite{KesselsJAP01v89}. Therefore, it is important 
to understand the production of SiH$_{3}$  and its role on the film 
deposition process.  We see 
from  equations (3) and (4) 
that  SiH$_{3}$$^{-}$ and SiH$_{3}$$^{+}$
 are produced in the fragmentation process 
after  attaching an electron  with SiH$_{4}$ and hydrogen atom with 
SiH$_{4}$$^{+}$.   This     
indicates   that the production of  SiH$_{3}$ can be improved  by   
controlling the electron and H content in PECVD plasma. 
It is interesting to compare these processes with the 
experimental investigation\cite{SandenJAP98v84} 
on argon/hydrogen expanding thermal
plasma. The investigation  
suggests that a significant amount of SiH$_{3}$ is produced 
by electron collision and hydrogen collision 
process.  

The structure and dipole moment of 
SiH$_{3}$ may play a role on improving a:Si-H film 
quality.  As we have mentioned, SiH$_{3}$ has low inversion 
barriers of 0.22 eV. This implies that SiH$_{3}$  
can transform to  planar structure from its ground state pyramidal 
structure by crossing the barrier. The barrier crossing can easily occur
in the hot plasma environment. When 
the cluster transforms into the planar structure, its dipole moment changes by 
a considerable amount. The dipole moment of  
pyramidal SiH$_{3}$ is 0.16D and that of planar SiH$_{3}$ is zero. Because 
of its zero dipole moment the planar SiH$_{3}$ does not get easily 
physisorbed on the film. This 
low physisorbtion behavior of planar SiH$_{3}$ 
cluster is similar to that of  CH$_{3}$ in 
hydrogenated amorphous carbon
 film deposition\cite{PerrinJVST98v16}. Since the planar 
SiH$_{3}$  has low physisorbtion property, it can 
 migrate on the surface until it  gets absorbed by a highly  reactive 
site which could be a  dangling bond, surface void or some 
defect.  The surface gets smoothened due to  the migration of   
planar SiH$_{3}$,  as a result the film quality improves.

\subsection{Formation of bigger clusters} 

We have shown in the silane fragmentation process that 
the product radicals are SiH$_{2}$, SiH$_{2}$$^{+}$, 
SiH$_{3}$$^{+}$ and SiH$_{3}$$^{-}$.  These radicals are 
reactive as they have unsaturated bonds and can react with 
silane and with each other forming  new 
species\cite{GallagherJAP05v98, GallagherJAP04v96, GallagherJAP02v91, 
GallagherPRE00v62, GallagherJAP00v87, GallagherJAP92v71b}.  The new species 
formed in the clustering process of   
 the radicals with silane are 

\begin{equation}
SiH_{2} + SiH_{4} \rightarrow Si_{2}H_{6} ( -2.29 e V ) 
\end{equation}
\begin{equation}
SiH_{2}^{+} + SiH_{4} \rightarrow Si_{2}H_{6}^{+} ( -1.68 eV ) 
\end{equation}
\begin{equation}
SiH_{3}^{+} + SiH_{4} \rightarrow Si_{2}H_{5}^{+} + H_{2} ( -0.33 eV )  
\end{equation}
\begin{equation}
SiH_{3}^{-} + SiH_{4} \rightarrow Si_{2}H_{5}^{-} + H_{2} ( -0.29 eV ) 
\end{equation}

As we see from the above equations, the interaction of SiH$_{2}$ with
SiH$_{4}$ resulting in Si$_{2}$H$_{6}$ is highly favored which agrees 
well with the experimental 
results\cite{GallagherJAP04v96, GallagherJAP92v71}. 

The species formed by clustering  of the radicals  with each other 
are

\begin{equation}
SiH_{2} + SiH_{2} \rightarrow Si_{2}H_{4} ( -2.63 eV ) 
\end{equation}
\begin{equation}
SiH_{2} + SiH_{2}^{+} \rightarrow Si_{2}H_{4}^{+} ( -3.71 eV ) 
\end{equation}
\begin{equation}
SiH_{2} + SiH_{3}^{+} \rightarrow Si_{2}H_{5}^{+} ( -2.90 eV ) 
\end{equation}
\begin{equation}
SiH_{2} + SiH_{3}^{-} \rightarrow Si_{2}H_{5}^{-}  ( -2.86 eV ) 
\end{equation}
\begin{equation}
SiH_{3}^{+} + SiH_{3}^{-} \rightarrow Si_{2}H_{6} ( -10.00 eV ) 
\end{equation}

The number in the parentheses is the cluster formation energy for 
the process, which is calculated by taking the total energy difference 
between the product and the reactant clusters. 

Except Si$_{2}$H$_{6}$ and Si$_{2}$H$_{6}$$^{+}$, other products formed in 
the above reactions 
undergo further clustering processes as they have unsaturated bonds. 
Since Si$_{2}$H$_{6}$$^{+}$ is saturated with hydrogen, it favors 
fragmentation.  Si$_{2}$H$_{6}$ is the immediate stable cluster formed 
by the reaction of 
silane fragmented  radicals. It is interesting to see that 
recent experimental  
observations\cite{GallagherJAP04v96, GallagherJAP92v71, 
DuanAPL01v78, SandenJAP98v84}
show sufficient  
amount of Si$_{2}$H$_{6}$ in silane decomposed PECVD plasma.

\section{Conclusions}

Using the  \emph{ab initio} electronic structure methods we have found 
the ground state structures of neutral and charged silane, disilane and 
their decomposed species.  Interestingly, in addition
to usual Si-H bond, hydrogen 
forms Si-H-Si and Si-H-H bonds in some of these clusters; it
forms Si-H-Si bridge bond in 
 Si$_{2}$H, Si$_{2}$H$_{2}$ and  
Si$_{2}$H$_{3}$ clusters and three center two electron  Si-H-H  bond 
in SiH$_{4}$$^{+}$ and SiH$_{5}$$^{+}$ clusters. We find that 
charging the clusters induces  structural 
modifications which mainly occur through Si-H bond 
orientations. We attribute   
these  structural modifications  to   
electrostatic repulsion between the non-bonded electrons and bonded 
electrons.

We find that Si-H bond  of Si$_{m}$H$_{n}$ cluster is shortened when 
positively charged  and  expanded when negatively 
charged. On the other hand  
Si-Si and Si-H-Si bonds of the cluster expand irrespective of 
the charged state. The bond length variations of 
Si-H and Si-Si bonds 
in the charged clusters are similar to that of charged SiH and Si$_{2}$ 
dimers.  

 Our vibrational analysis 
shows that frequencies of Si-H, Si-Si, Si-H-Si stretch modes 
show significant changes when the clusters are charged. These frequency 
shifts are 
consistent with their bond length variations.  

We have shown that the low energy pathways of fragmenting  
neutral and charged silane are associated with  
SiH$_{2}$, SiH$_{2}$$^{+}$, SiH$_{3}$$^{+}$ and SiH$_{3}$$^{-}$ 
products. The fragmentation results 
indicate that  
the production of  SiH$_{3}$  can be controlled by electron and
H content in the PECVD plasma. We speculate that  
the experimentally observed\cite{KesselsJAP01v89} improvement in  
{\it{a}}-S:H film quality by SiH$_{3}$ deposition is  
due to the surface migration of planar SiH$_{3}$.  The investigation on 
the clustering process shows that Si$_{2}$H$_{6}$ is 
the immediate stable cluster formed by the silane decomposed 
radicals which is  consistent with the experimental 
observations\cite{GallagherJAP04v96, GallagherJAP92v71, 
DuanAPL01v78, SandenJAP98v84}

\begin{center}
\textbf{Acknowledgment}
\end{center}

 It is a pleasure to thank Profs. S. C. Agarwal and M. K. Harbola 
for helpful discussions
and comments. This work was supported by the Department of Science
and Technology, New Delhi via project No. SP/S2/M-51/96.

\newpage
Table. I. Our calculated ionization potentials ( IP )  of Si$_{m}$H$_{n}$ 
clusters using Gaussian98 package. The experimental  
values  (a)\cite{BerkowitzJCP87v86},  
(b)\cite{ItohJCP86v85} and  (c)\cite{RobertsonJAP86v59} are given 
in the last column. 

\begin{longtable}[c]{|p{1.0cm}|p{1.8cm}|p{3.2cm}|}\hline

	&{\centering {\sf Gaussian98}}
	&{\centering {\sf Experiment}}
\\\hline
	
	&{\centering {\sf IP ( eV ) }}
	&{\centering {\sf IP ( eV ) }}
\\\hline
	 {\centering {\sf Si}}
	&{\centering {\sf 8.1}}
	&{\centering {\sf 8.2(a)}}
\\\hline
	 {\centering {\sf SiH}}
	&{\centering {\sf 8.0}}
	&{\centering {\sf 7.9(a)}}
\\\hline
	 {\centering {\sf SiH$_{2}$}}
	&{\centering {\sf 9.0}}
	&{\centering {\sf 9(a)}}
\\\hline
	 {\centering {\sf SiH$_{3}$}}
	&{\centering {\sf 8.1}}
	&{\centering {\sf 8(a)}}
\\\hline
	 {\centering {\sf SiH$_{4}$}}
	&{\centering {\sf 11.0}}
	&{\centering {\sf 11(a), 12.4(b)}}
\\\hline
	 {\centering {\sf Si$_{2}$}}
	&{\centering {\sf 7.8}}
	&{\centering {\sf 7.7(c)}}
\\\hline
	 {\centering {\sf Si$_{2}$H}}
	&{\centering {\sf 8.4}}
	&{\centering {\sf 7.9(c)}}
\\\hline
	 {\centering {\sf Si$_{2}$H$_{2}$}}
	&{\centering {\sf 8.0}}
	&{\centering {\sf 7.7(c)}}
\\\hline
	 {\centering {\sf Si$_{2}$H$_{3}$}}
	&{\centering {\sf 7.8}}
	&{\centering {\sf 7.8(c)}}
\\\hline
	 {\centering {\sf Si$_{2}$H$_{4}$}}
	&{\centering {\sf 7.9}}
	&{\centering {\sf 7.6(c)}}
\\\hline
	 {\centering {\sf Si$_{2}$H$_{5}$}}
	&{\centering {\sf 7.7}}
	&{\centering {\sf 7.7(c)}}
\\\hline
	 {\centering {\sf Si$_{2}$H$_{6}$}}
	&{\centering {\sf 9.6}}
	&{\centering {\sf 7.9(c)}}
\\\hline
\end{longtable}

\newpage

Table. II The calculated vibrational frequencies $\nu$ ( cm$^{-1}$ ) and 
force constants k ( mDyne/$\AA$ ) of neutral and charged 
SiH$_{m}$( m=1-4) and 
also charged SiH$_{5}$  clusters using B3LYP functional employing 
6-311g** basis set. Important modes such as stretch, breathing etc., 
have been identified. 

\begin{longtable}[c]{|p{0.5cm}|p{3cm}|p{1.6cm}|p{1.6cm}|p{1.6cm}|
p{1.6cm}|p{1.6cm}|p{1.6cm}|}\hline
&
&
\multicolumn{1}{c}{SiH}
&
&
\multicolumn{1}{c}{SiH$^{+}$}
&
&
\multicolumn{1}{c}{SiH$^{-}$}
&
\\\hline
        &{\centering {\sf mode}}
	&{\centering {\sf\ \ \  $\nu$ }}
	&{\centering {\sf\ \ \ k }}
	&{\centering {\sf\ \ \ $\nu$ }}
	&{\centering {\sf\ \ \ k }}
	&{\centering {\sf\ \ \ $\nu$ }}
	&{\centering {\sf\ \ \ k }}
\\\hline 
	 {\centering {\sf 1}}
	&{\centering {\sf Si-H stretch}}
	&{\centering {\sf 2013}}
	&{\centering {\sf 2.49}}
	&{\centering {\sf 2137}}
	&{\centering {\sf 2.81}}
	&{\centering {\sf 1827}}
	&{\centering {\sf 2.0512}}
\\\hline
\hline
&	
&
\multicolumn{1}{c}{SiH$_{2}$}
&
&
\multicolumn{1}{c}{SiH$_{2}$$^{+}$}
&
&
\multicolumn{1}{c}{SiH$_{2}$$^{-}$}
&
\\\hline
	 {\centering {\sf 1}}
	&{\centering {\sf H-Si-H breathing}}
	&{\centering {\sf 1024}}
	&{\centering {\sf 0.64}}
	&{\centering {\sf 907}}
	&{\centering {\sf 0.51}}
	&{\centering {\sf 967}}
	&{\centering {\sf 0.58}}
\\\hline
	 {\centering {\sf 2}}
	&{\centering {\sf Si-H stretch }}
	&{\centering {\sf 2036}}
	&{\centering {\sf 2.55}}
	&{\centering {\sf 2140}}
	&{\centering {\sf 2.76}}
	&{\centering {\sf 1843}}
	&{\centering {\sf 2.09}}
\\\hline
	 {\centering {\sf 3}}
	&{\centering {\sf Si-H stretch }}
	&{\centering {\sf 2038}}
	&{\centering {\sf 2.55}}
	&{\centering {\sf 2225}}
	&{\centering {\sf 3.09}}
	&{\centering {\sf 1852}}
	&{\centering {\sf 2.10}}
\\\hline
\hline
&	
&
\multicolumn{1}{c}{SiH$_{3}$}
&
&
\multicolumn{1}{c}{SiH$_{3}$$^{+}$}
&
&
\multicolumn{1}{c}{SiH$_{3}$$^{-}$}
&
\\\hline
	 {\centering {\sf 1}}
	&{\centering {\sf Umbrella}}
	&{\centering {\sf 760}}
	&{\centering {\sf 0.37}}
	&{\centering {\sf 839}}
	&{\centering {\sf 1.11}}
	&{\centering {\sf 894}}
	&{\centering {\sf 0.51}}
\\\hline
	 {\centering {\sf 2}}
	&{\centering {\sf }}
	&{\centering {\sf 936}}
	&{\centering {\sf 0.54}}
	&{\centering {\sf 940}}
	&{\centering {\sf 1.11}}
	&{\centering {\sf 972}}
	&{\centering {\sf 0.51}}
\\\hline
	 {\centering {\sf 3}}
	&
	&{\centering {\sf 936}}
	&{\centering {\sf 0.54}}
	&{\centering {\sf 940}}
	&{\centering {\sf 1.05}}
	&{\centering {\sf 972}}
	&{\centering {\sf 0.57}}
\\\hline
	 {\centering {\sf 4}}
	&{\centering {\sf Si-H stretch }}
	&{\centering {\sf 2192}}
	&{\centering {\sf 2.87}}
	&{\centering {\sf 2274}}
	&{\centering {\sf 3.07}}
	&{\centering {\sf 1885}}
	&{\centering {\sf 2.19}}
\\\hline
	 {\centering {\sf 5}}
	&{\centering {\sf Si-H stretch }}
	&{\centering {\sf 2233}}
	&{\centering {\sf 3.10}}
	&{\centering {\sf 2356}}
	&{\centering {\sf 3.47}}
	&{\centering {\sf 1885}}
	&{\centering {\sf 2.19}}
\\\hline
	 {\centering {\sf 6}}
	&{\centering {\sf Si-H stretch }}
	&{\centering {\sf 2233}}
	&{\centering {\sf 3.10}}
	&{\centering {\sf 2357}}
	&{\centering {\sf 3.47}}
	&{\centering {\sf 1899}}
	&{\centering {\sf 2.19}}
\\\hline
\hline
&
&
\multicolumn{1}{c}{SiH$_{4}$}
&
&
\multicolumn{1}{c}{SiH$_{4}$$^{+}$}
&
&
\multicolumn{1}{c}{SiH$_{4}$$^{-}$}
&
\\\hline
	 {\centering {\sf 1}}
	&
	&{\centering {\sf 922}}
	&{\centering {\sf 0.55}}
	&{\centering {\sf 244}}
	&{\centering {\sf 0.04}}
	&{\centering {\sf 712}}
	&{\centering {\sf 0.04}}
\\\hline
	 {\centering {\sf 2}}
	&
	&{\centering {\sf 922}}
	&{\centering {\sf 0.55}}
	&{\centering {\sf 622}}
	&{\centering {\sf 0.26}}
	&{\centering {\sf 796}}
	&{\centering {\sf 0.39}}
\\\hline
	 {\centering {\sf 3}}
	&
	&{\centering {\sf 922}}
	&{\centering {\sf 0.55}}
	&{\centering {\sf 675}}
	&{\centering {\sf 0.28}}
	&{\centering {\sf 814}}
	&{\centering {\sf 0.42}}
\\\hline
	 {\centering {\sf 4}}
	&
	&{\centering {\sf 980}}
	&{\centering {\sf 0.57}}
	&{\centering {\sf 751}}
	&{\centering {\sf 0.34}}
	&{\centering {\sf 992}}
	&{\centering {\sf 0.63}}
\\\hline
	 {\centering {\sf 5}}
	&
	&{\centering {\sf 980}}
	&{\centering {\sf 0.57}}
	&{\centering {\sf 874}}
	&{\centering {\sf 0.48}}
	&{\centering {\sf 1038}}
	&{\centering {\sf 0.64}}
\\\hline
	 {\centering {\sf 6}}
	&
	&{\centering {\sf 2235}}
	&{\centering {\sf 2.97}}
	&{\centering {\sf 1043}}
	&{\centering {\sf 0.65}}
	&{\centering {\sf 1341}}
	&{\centering {\sf 1.11}}
\\\hline
	 {\centering {\sf 7}}
	&
	&{\centering {\sf 2243}}
	&{\centering {\sf 3.12}}
	&{\centering {\sf 2216}}
	&{\centering {\sf 2.96}}
	&{\centering {\sf 1448}}
	&{\centering {\sf 1.25}}
\\\hline
	 {\centering {\sf 8}}
	&
	&{\centering {\sf 2244}}
	&{\centering {\sf 3.12}}
	&{\centering {\sf 2292}}
	&{\centering {\sf 3.29}}
	&{\centering {\sf 2007}}
	&{\centering {\sf 2.47}}
\\\hline
	 {\centering {\sf 9}}
	&
	&{\centering {\sf 2244}}
	&{\centering {\sf 3.12}}
	&{\centering {\sf 3794}}
	&{\centering {\sf 8.55}}
	&{\centering {\sf 2008}}
	&{\centering {\sf 2.46}}
\\\hline
&	
&
\multicolumn{1}{c}{SiH$_{5}$$^{+}$}
&
&
\multicolumn{1}{c}{SiH$_{5}$$^{-}$}
&
&
&
\\\hline
	 {\centering {\sf 1}}
        &
	&{\centering {\sf 63}}
	&{\centering {\sf 0.00}}
	&{\centering {\sf 513}}
	&{\centering {\sf 0.16}}
	&
        &
\\\hline
	{\centering {\sf 2}}
        &
	&{\centering {\sf 636}}
	&{\centering {\sf 0.25}}
	&{\centering {\sf 516}}
	&{\centering {\sf 0.16}}
	&
	&
\\\hline
 	 {\centering {\sf 3}}
&
	&{\centering {\sf 656}}
	&{\centering {\sf 0.26}}
	&{\centering {\sf 992}}
	&{\centering {\sf 0.66}}
	&
	&
\\\hline
	 {\centering {\sf 4}}
&
	&{\centering {\sf 745}}
	&{\centering {\sf 0.37}}
	&{\centering {\sf 1041}}
	&{\centering {\sf 0.71}}
	&
	&
\\\hline
	 {\centering {\sf 5}}
&
	&{\centering {\sf 869}}
	&{\centering {\sf 0.46}}
	&{\centering {\sf 1041}}
	&{\centering {\sf 0.71}}
	&
	&
\\\hline
	 {\centering {\sf 6}}
&
	&{\centering {\sf 923}}
	&{\centering {\sf 0.53}}
	&{\centering {\sf 1204}}
	&{\centering {\sf 0.86}}
	&
	&
\\\hline
	 {\centering {\sf 7}}
&
	&{\centering {\sf 926}}
	&{\centering {\sf 0.53}}
	&{\centering {\sf 1205}}
	&{\centering {\sf 0.86}}
	&
	&
\\\hline
	 {\centering {\sf 8}}
&
	&{\centering {\sf 1046}}
	&{\centering {\sf 0.65}}
	&{\centering {\sf 1354}}
	&{\centering {\sf 1.09}}
	&
	&
\\\hline
	 {\centering {\sf 9}}
&
	&{\centering {\sf 2289}}
	&{\centering {\sf 3.12}}
	&{\centering {\sf 1506}}
	&{\centering {\sf 1.38}}
	&
	&
\\\hline
	 {\centering {\sf 10}}
&
	&{\centering {\sf 2350}}
	&{\centering {\sf 3.45}}
	&{\centering {\sf 1924}}
	&{\centering {\sf 2.29}}
	&
	&
\\\hline
	 {\centering {\sf 11}}
&
	&{\centering {\sf 2360}}
	&{\centering {\sf 3.48}}
	&{\centering {\sf 1924}}
	&{\centering {\sf 2.29}}
	&
	&
\\\hline
	 {\centering {\sf 12}}
&
	&{\centering {\sf 3955}}
	&{\centering {\sf 9.29}}
	&{\centering {\sf 1955}}
	&{\centering {\sf 2.27}}
	&
	&
\\\hline
\end{longtable}

\newpage
Table. III The calculated vibrational frequencies $\nu$ ( cm$^{-1}$ ) and
force constants k ( mDyne/$\AA$ ) of neutral and 
charged Si$_{2}$H$_{n}$( n=1-4) 
clusters using B3LYP functional employing
6-311g** basis set. Si-H-Si ( $\parallel$ ) mode corresponds to the motion of 
H atom parallel to Si-Si bond of the cluster 
and Si-H-Si ( $\perp$ ) corresponds to 
the motion of H atom perpendicular to Si-Si bond. 

\begin{longtable}[c]{|p{0.5cm}|p{3cm}|p{1.6cm}|p{1.6cm}|p{1.6cm}|
p{1.6cm}|p{1.6cm}|p{1.6cm}|}\hline
&
&
\multicolumn{1}{c}{Si$_{2}$}
&
&
\multicolumn{1}{c}{Si$_{2}$$^{+}$}
&
&
\multicolumn{1}{c}{Si$_{2}$$^{-}$}
&
\\\hline
	
	&{\centering {\sf mode}}
	&{\centering {\sf $\nu$}}
	&{\centering {\sf k}}
	&{\centering {\sf $\nu$}}
	&{\centering {\sf k}}
	&{\centering {\sf $\nu$}}
	&{\centering {\sf k}}
\\\hline
	 {\centering {\sf 1}}
	&{\centering {\sf Si-Si stretch}}
	&{\centering {\sf 600}}
	&{\centering {\sf 5.94}}
	&{\centering {\sf 546}}
	&{\centering {\sf 4.92}}
	&{\centering {\sf 586}}
	&{\centering {\sf 5.65}}
\\\hline
&
&
\multicolumn{1}{c}{Si$_{2}$H}
&
&
\multicolumn{1}{c}{Si$_{2}$H$^{+}$}
&
&
\multicolumn{1}{c}{Si$_{2}$H$^{-}$}
&
\\\hline
	 {\centering {\sf 1}}
	&{\centering {\sf Si-Si stretch}}
	&{\centering {\sf 543}}
	&{\centering {\sf 4.02}}
	&{\centering {\sf 407}}
	&{\centering {\sf 2.22}}
	&{\centering {\sf 556}}
	&{\centering {\sf 4.29}}
\\\hline
	 {\centering {\sf 2}}
	&{\centering {\sf Si-H-Si ( $\parallel$ )}}
	&{\centering {\sf 1108}}
	&{\centering {\sf 0.76}}
	&{\centering {\sf 890}}
	&{\centering {\sf 0.49}}
	&{\centering {\sf 1104}}
	&{\centering {\sf 0.75}}
\\\hline
	 {\centering {\sf 3}}
	&{\centering {\sf Si-H-Si ( $\perp$ )}}
	&{\centering {\sf 1565}}
	&{\centering {\sf 1.49}}
	&{\centering {\sf 1384}}
	&{\centering {\sf 1.17}}
	&{\centering {\sf 1479}}
	&{\centering {\sf 1.33}}
\\\hline
&
&
\multicolumn{1}{c}{Si$_{2}$H$_{2}$}
&
&
\multicolumn{1}{c}{Si$_{2}$H$_{2}$$^{+}$}
&
&
\multicolumn{1}{c}{Si$_{2}$H$_{2}$$^{-}$}
&
\\\hline
	 {\centering {\sf 1}}
	&{\centering {\sf Si-Si stretch}}
	&{\centering {\sf 522}}
	&{\centering {\sf 1.93}}
	&{\centering {\sf 395}}
	&{\centering {\sf 1.00}}
	&{\centering {\sf 346}}
	&{\centering {\sf 0.40}}
\\\hline
	 {\centering {\sf 2}}
	&{\centering {\sf }}
	&{\centering {\sf 963}}
	&{\centering {\sf 0.59}}
	&{\centering {\sf 993}}
	&{\centering {\sf 0.63}}
	&{\centering {\sf 782}}
	&{\centering {\sf 0.44}}
\\\hline
	 {\centering {\sf 3}}
	&{\centering {\sf Si-H-Si ($\parallel$)}}
	&{\centering {\sf 1083}}
	&{\centering {\sf 0.72}}
	&{\centering {\sf 1054}}
	&{\centering {\sf 0.68}}
	&{\centering {\sf 948}}
	&{\centering {\sf 0.55}}
\\\hline
	 {\centering {\sf 4}}
	&{\centering {\sf Si-H-Si ( $\parallel$ )}}
	&{\centering {\sf 1175}}
	&{\centering {\sf 0.86}}
	&{\centering {\sf 1171}}
	&{\centering {\sf 0.86}}
	&{\centering {\sf 954}}
	&{\centering {\sf 0.56}}
\\\hline
	 {\centering {\sf 5}}
	&{\centering {\sf Si-H-Si ( $\perp$ )}}
	&{\centering {\sf 1534}}
	&{\centering {\sf 1.43}}
	&{\centering {\sf 1546}}
	&{\centering {\sf 1.45}}
	&{\centering {\sf 1418}}
	&{\centering {\sf 1.22}}
\\\hline
	 {\centering {\sf 6}}
	&{\centering {\sf Si-H-Si ( $\perp$ ) }}
	&{\centering {\sf 1616}}
	&{\centering {\sf 1.57}}
	&{\centering {\sf 1626}}
	&{\centering {\sf 1.59}}
	&{\centering {\sf 1550}}
	&{\centering {\sf 1.44}}
\\\hline
&
&
\multicolumn{1}{c}{Si$_{2}$H$_{3}$}
&
&
\multicolumn{1}{c}{Si$_{2}$H$_{3}$$^{+}$}
&
&
\multicolumn{1}{c}{Si$_{2}$H$_{3}$$^{-}$}
&

\\\hline
	 {\centering {\sf 1}}
	&
	&{\centering {\sf 358.}}
	&{\centering {\sf 0.07}}
	&{\centering {\sf 198}}
	&{\centering {\sf 0.02}}
	&{\centering {\sf 496}}
	&{\centering {\sf 0.15}}
\\\hline
	 {\centering {\sf 2}}
	&{\centering {\sf Si-Si stretch}}
	&{\centering {\sf 446.}}
	&{\centering {\sf 1.57}}
	&{\centering {\sf 442}}
	&{\centering {\sf 0.38}}
	&{\centering {\sf 415}}
	&{\centering {\sf 1.53}}
\\\hline
	 {\centering {\sf 3}}
	&
	&{\centering {\sf 647}}
	&{\centering {\sf 0.26}}
	&{\centering {\sf 575}}
	&{\centering {\sf 0.27}}
	&{\centering {\sf 688}}
	&{\centering {\sf 0.29}}
\\\hline
	 {\centering {\sf 4}}
	&
	&{\centering {\sf 739.}}
	&{\centering {\sf 0.35}}
	&{\centering {\sf 637}}
	&{\centering {\sf 0.27}}
	&{\centering {\sf 788}}
	&{\centering {\sf 0.39}}
\\\hline
	 {\centering {\sf 5}}
	&
	&{\centering {\sf 741.}}
	&{\centering {\sf 0.34}}
	&{\centering {\sf 772}}
	&{\centering {\sf 0.38}}
	&{\centering {\sf 802}}
	&{\centering {\sf 0.41}}
\\\hline
	 {\centering {\sf 6}}
	&{\centering {\sf Si-H-Si ( $\parallel$ )}}
	&{\centering {\sf 1090}}
	&{\centering {\sf 0.71}}
	&{\centering {\sf 1082}}
	&{\centering {\sf 0.70}}
	&{\centering {\sf 1118}}
	&{\centering {\sf 0.75}}
\\\hline
	 {\centering {\sf 7}}
	&{\centering {\sf Si-H-Si ( $\perp$ ) }}
	&{\centering {\sf 1497.}}
	&{\centering {\sf 1.37}}
	&{\centering {\sf 1606}}
	&{\centering {\sf 1.57}}
	&{\centering {\sf 1375}}
	&{\centering {\sf 1.16}}
\\\hline
	 {\centering {\sf 8}}
	&{\centering {\sf Si-H stretch }}
	&{\centering {\sf 2082}}
	&{\centering {\sf 2.66}}
	&{\centering {\sf 2192}}
	&{\centering {\sf 2.96}}
	&{\centering {\sf 1888}}
	&{\centering {\sf 2.19}}
\\\hline
	 {\centering {\sf 9}}
	&{\centering {\sf Si-H stretch }}
	&{\centering {\sf 2089.}}
	&{\centering {\sf 2.68}}
	&{\centering {\sf 2199}}
	&{\centering {\sf 2.97}}
	&{\centering {\sf 1908}}
	&{\centering {\sf 2.24}}
\\\hline
&
&
\multicolumn{1}{c}{Si$_{2}$H$_{4}$}
&
&
\multicolumn{1}{c}{Si$_{2}$H$_{4}$$^{+}$}
&
&
\multicolumn{1}{c}{Si$_{2}$H$_{4}$$^{-}$}
&
\\\hline
	 {\centering {\sf 1}}
	&
	&{\centering {\sf 323}}
	&{\centering {\sf 0.09}}
	&{\centering {\sf 336}}
	&{\centering {\sf 0.08}}
	&{\centering {\sf 198}}
	&{\centering {\sf 0.02}}
\\\hline
	 {\centering {\sf 2}}
	&
	&{\centering {\sf 348}}
	&{\centering {\sf 0.07}}
	&{\centering {\sf 345}}
	&{\centering {\sf 0.07}}
	&{\centering {\sf 360}}
	&{\centering {\sf 0.08}}
\\\hline
	 {\centering {\sf 3}}
	&
	&{\centering {\sf 448}}
	&{\centering {\sf 0.13}}
	&{\centering {\sf 370}}
	&{\centering {\sf 0.08}}
	&{\centering {\sf 458}}
	&{\centering {\sf 0.12}}
\\\hline
	 {\centering {\sf 4}}
	&
	&{\centering {\sf 525}}
	&{\centering {\sf 0.16}}
	&{\centering {\sf 524}}
	&{\centering {\sf 0.17}}
	&{\centering {\sf 676}}
	&{\centering {\sf 0.31}}
\\\hline
	 {\centering {\sf 5}}
	&{\centering {\sf Si-Si stretch}}
	&{\centering {\sf 563}}
	&{\centering {\sf 0.78}}
	&{\centering {\sf 505}}
	&{\centering {\sf 1.30}}
	&{\centering {\sf 390}}
	&{\centering {\sf 0.95}}
\\\hline
	 {\centering {\sf 6}}
	&
	&{\centering {\sf 616}}
	&{\centering {\sf 0.27}}
	&{\centering {\sf 607}}
	&{\centering {\sf 0.26}}
	&{\centering {\sf 677}}
	&{\centering {\sf 0.30}}
\\\hline
	 {\centering {\sf 7}}
	&
	&{\centering {\sf 919}}
	&{\centering {\sf 0.53}}
	&{\centering {\sf 853}}
	&{\centering {\sf 0.45}}
	&{\centering {\sf 936}}
	&{\centering {\sf 0.54}}
\\\hline
	 {\centering {\sf 8}}
	&
	&{\centering {\sf 956}}
	&{\centering {\sf 0.58}}
	&{\centering {\sf 942}}
	&{\centering {\sf 0.56}}
	&{\centering {\sf 943}}
	&{\centering {\sf 0.55}}
\\\hline
	 {\centering {\sf 9}}
	&{\centering {\sf Si-H stretch}}
	&{\centering {\sf 2224}}
	&{\centering {\sf 3.00}}
	&{\centering {\sf 2279}}
	&{\centering {\sf 3.14}}
	&{\centering {\sf 1998}}
	&{\centering {\sf 2.43}}
\\\hline
	 {\centering {\sf 10}}
	&{\centering {\sf Si-H stretch }}
	&{\centering {\sf 2228}}
	&{\centering {\sf 3.01}}
	&{\centering {\sf 2281}}
	&{\centering {\sf 3.14}}
	&{\centering {\sf 2000}}
	&{\centering {\sf 2.47}}
\\\hline
	 {\centering {\sf 11}}
	&{\centering {\sf Si-H stretch }}
	&{\centering {\sf 2245}}
	&{\centering {\sf 3.13}}
	&{\centering {\sf 2326}}
	&{\centering {\sf 3.38}}
	&{\centering {\sf 2009}}
	&{\centering {\sf 2.46}}
\\\hline
	 {\centering {\sf 12}}
	&{\centering {\sf Si-H stretch }}
	&{\centering {\sf 2257}}
	&{\centering {\sf 3.17}}
	&{\centering {\sf 2334}}
	&{\centering {\sf 3.40}}
	&{\centering {\sf 2019}}
	&{\centering {\sf 2.52}}
\\\hline
&
&
\multicolumn{1}{c}{Si$_{2}$H$_{5}$}
&
&
\multicolumn{1}{c}{Si$_{2}$H$_{5}$$^{+}$}
&
&
\multicolumn{1}{c}{Si$_{2}$H$_{5}$$^{-}$}
&
\\\hline
	 {\centering {\sf 1}}
	&
	&{\centering {\sf 128}}
	&{\centering {\sf 0.01}}
	&{\centering {\sf 40}}
	&{\centering {\sf 0.00}}
	&{\centering {\sf 175}}
	&{\centering {\sf 0.02}}
\\\hline
	 {\centering {\sf 2}}
	&
	&{\centering {\sf 390}}
	&{\centering {\sf 0.09}}
	&{\centering {\sf 340}}
	&{\centering {\sf 0.11}}
	&{\centering {\sf 366}}
	&{\centering {\sf 0.70}}
\\\hline
	 {\centering {\sf 3}}
	&
	&{\centering {\sf 406}}
	&{\centering {\sf 0.12}}
	&{\centering {\sf 362}}
	&{\centering {\sf 0.08}}
	&{\centering {\sf 390}}
	&{\centering {\sf 0.09}}
\\\hline
	 {\centering {\sf 4}}
	&{\centering {\sf Si-Si stretch}}
	&{\centering {\sf 424}}
	&{\centering {\sf 0.41}}
	&{\centering {\sf 397}}
	&{\centering {\sf 0.21}}
	&{\centering {\sf 398}}
	&{\centering {\sf 0.10}}
\\\hline
	 {\centering {\sf 5}}
	&
	&{\centering {\sf 597}}
	&{\centering {\sf 0.25}}
	&{\centering {\sf 608}}
	&{\centering {\sf 0.26}}
	&{\centering {\sf 676}}
	&{\centering {\sf 0.32}}
\\\hline
	 {\centering {\sf 6}}
	&
	&{\centering {\sf 637}}
	&{\centering {\sf 0.28}}
	&{\centering {\sf 676}}
	&{\centering {\sf 0.32}}
	&{\centering {\sf 696}}
	&{\centering {\sf 0.32}}
\\\hline
	 {\centering {\sf 7}}
	&
	&{\centering {\sf 876}}
	&{\centering {\sf 0.49}}
	&{\centering {\sf 817}}
	&{\centering {\sf 0.43}}
	&{\centering {\sf 923}}
	&{\centering {\sf 0.55}}
\\\hline
	 {\centering {\sf 8}}
	&
	&{\centering {\sf 936}}
	&{\centering {\sf 0.55}}
	&{\centering {\sf 910}}
	&{\centering {\sf 0.50}}
	&{\centering {\sf 939}}
	&{\centering {\sf 0.54}}
\\\hline
	 {\centering {\sf 9}}
	&
	&{\centering {\sf 949}}
	&{\centering {\sf 0.56}}
	&{\centering {\sf 921}}
	&{\centering {\sf 0.52}}
	&{\centering {\sf 970}}
	&{\centering {\sf 0.58}}
\\\hline
	 {\centering {\sf 10}}
	&
	&{\centering {\sf 951}}
	&{\centering {\sf 0.55}}
	&{\centering {\sf 976}}
	&{\centering {\sf 0.60}}
	&{\centering {\sf 974}}
	&{\centering {\sf 0.58}}
\\\hline
	 {\centering {\sf 11}}
	&{\centering {\sf Si-H stretch}}
	&{\centering {\sf 2188}}
	&{\centering {\sf 2.90}}
	&{\centering {\sf 2235}}
	&{\centering {\sf 3.01}}
	&{\centering {\sf 1932}}
	&{\centering {\sf 2.30}}
\\\hline
	 {\centering {\sf 12}}
	&{\centering {\sf Si-H stretch}}
	&{\centering {\sf 2199}}
	&{\centering {\sf 2.95}}
	&{\centering {\sf 2252}}
	&{\centering {\sf 3.08}}
	&{\centering {\sf 1935}}
	&{\centering {\sf 2.28}}
\\\hline
	 {\centering {\sf 13}}
	&{\centering {\sf Si-H stretch}}
	&{\centering {\sf 2214}}
	&{\centering {\sf 3.04}}
	&{\centering {\sf 2280}}
	&{\centering {\sf 3.19}}
	&{\centering {\sf 2040}}
	&{\centering {\sf 2.56}}
\\\hline
	 {\centering {\sf 14}}
	&{\centering {\sf Si-H stretch}}
	&{\centering {\sf 2223}}
	&{\centering {\sf 3.01}}
	&{\centering {\sf 2291}}
	&{\centering {\sf 3.26}}
	&{\centering {\sf 2072}}
	&{\centering {\sf 2.65}}
\\\hline
	 {\centering {\sf 15}}
	&{\centering {\sf Si-H stretch}}
	&{\centering {\sf 2231}}
	&{\centering {\sf 3.09}}
	&{\centering {\sf 2304}}
	&{\centering {\sf 3.30}}
	&{\centering {\sf 2083}}
	&{\centering {\sf 2.63}}
\\\hline
&
&
\multicolumn{1}{c}{Si$_{2}$H$_{6}$}
&
&
\multicolumn{1}{c}{Si$_{2}$H$_{6}$$^{+}$}
&
&
\multicolumn{1}{c}{Si$_{2}$H$_{6}$$^{-}$}
&
\\\hline
	 {\centering {\sf 1}}
	&
	&{\centering {\sf 131}}
	&{\centering {\sf 0.01}}
	&{\centering {\sf 105}}
	&{\centering {\sf 0.01}}
	&{\centering {\sf 162}}
	&{\centering {\sf 0.02}}
\\\hline
	 {\centering {\sf 2}}
	&
	&{\centering {\sf 380}}
	&{\centering {\sf 0.09}}
	&{\centering {\sf 208}}
	&{\centering {\sf 0.22}}
	&{\centering {\sf 211}}
	&{\centering {\sf 0.23}}
\\\hline
	 {\centering {\sf 3}}
	&
	&{\centering {\sf 380}}
	&{\centering {\sf 0.09}}
	&{\centering {\sf 265}}
	&{\centering {\sf 0.04}}
	&{\centering {\sf 250}}
	&{\centering {\sf 0.04}}
\\\hline
	 {\centering {\sf 4}}
	&{\centering {\sf Si-Si stretch}}
	&{\centering {\sf 423}}
	&{\centering {\sf 0.75}}
	&{\centering {\sf 267}}
	&{\centering {\sf 0.04}}
	&{\centering {\sf 285}}
	&{\centering {\sf 0.05}}
\\\hline
	 {\centering {\sf 5}}
	&
	&{\centering {\sf 636}}
	&{\centering {\sf 0.29}}
	&{\centering {\sf 328}}
	&{\centering {\sf 0.07}}
	&{\centering {\sf 498}}
	&{\centering {\sf 0.18}}
\\\hline
	 {\centering {\sf 6}}
	&
	&{\centering {\sf 636}}
	&{\centering {\sf 0.29}}
	&{\centering {\sf 328}}
	&{\centering {\sf 0.07}}
	&{\centering {\sf 562}}
	&{\centering {\sf 0.22}}
\\\hline
	 {\centering {\sf 7}}
	&
	&{\centering {\sf 855}}
	&{\centering {\sf 0.47}}
	&{\centering {\sf 723}}
	&{\centering {\sf 0.34}}
	&{\centering {\sf 855}}
	&{\centering {\sf 0.47}}
\\\hline
	 {\centering {\sf 8}}
	&
	&{\centering {\sf 928}}
	&{\centering {\sf 0.56}}
	&{\centering {\sf 789}}
	&{\centering {\sf 0.40}}
	&{\centering {\sf 884}}
	&{\centering {\sf 0.49}}
\\\hline
	 {\centering {\sf 9}}
	&
	&{\centering {\sf 945}}
	&{\centering {\sf 0.54}}
	&{\centering {\sf 918}}
	&{\centering {\sf 0.52}}
	&{\centering {\sf 932}}
	&{\centering {\sf 0.53}}
\\\hline
	 {\centering {\sf 10}}
	&
	&{\centering {\sf 945}}
	&{\centering {\sf 0.54}}
	&{\centering {\sf 918}}
	&{\centering {\sf 0.52}}
	&{\centering {\sf 938}}
	&{\centering {\sf 0.53}}
\\\hline
	 {\centering {\sf 11}}
	&
	&{\centering {\sf 959}}
	&{\centering {\sf 0.56}}
	&{\centering {\sf 928}}
	&{\centering {\sf 0.53}}
	&{\centering {\sf 962}}
	&{\centering {\sf 0.58}}
\\\hline
	 {\centering {\sf 12}}
	&
	&{\centering {\sf 959}}
	&{\centering {\sf 0.56}}
	&{\centering {\sf 928}}
	&{\centering {\sf 0.53}}
	&{\centering {\sf 966}}
	&{\centering {\sf 0.56}}
\\\hline
	 {\centering {\sf 13}}
	&{\centering {\sf Si-H stretch}}
	&{\centering {\sf 2209}}
	&{\centering {\sf 2.93}}
	&{\centering {\sf 2256}}
	&{\centering {\sf 3.03}}
	&{\centering {\sf 1765}}
	&{\centering {\sf 1.91}}
\\\hline
	 {\centering {\sf 14}}
	&{\centering {\sf Si-H stretch}}
	&{\centering {\sf 2217}}
	&{\centering {\sf 2.99}}
	&{\centering {\sf 2258}}
	&{\centering {\sf 3.03}}
	&{\centering {\sf 1826}}
	&{\centering {\sf 2.03}}
\\\hline
	 {\centering {\sf 15}}
	&{\centering {\sf Si-H stretch}}
	&{\centering {\sf 2218}}
	&{\centering {\sf 3.02}}
	&{\centering {\sf 2314}}
	&{\centering {\sf 3.33}}
	&{\centering {\sf 2058}}
	&{\centering {\sf 2.62}}
\\\hline
	 {\centering {\sf 16}}
	&{\centering {\sf Si-H stretch}}
	&{\centering {\sf 2219}}
	&{\centering {\sf 3.03}}
	&{\centering {\sf 2315}}
	&{\centering {\sf 3.34}}
	&{\centering {\sf 2061}}
	&{\centering {\sf 2.58}}
\\\hline
	 {\centering {\sf 17}}
	&{\centering {\sf Si-H stretch}}
	&{\centering {\sf 2227}}
	&{\centering {\sf 3.08}}
	&{\centering {\sf 2321}}
	&{\centering {\sf 3.36}}
	&{\centering {\sf 2069}}
	&{\centering {\sf 2.60}}
\\\hline
	 {\centering {\sf 18}}
	&{\centering {\sf Si-H stretch}}
	&{\centering {\sf 2229}}
	&{\centering {\sf 3.08}}
	&{\centering {\sf 2322}}
	&{\centering {\sf 3.36}}
	&{\centering {\sf 2076}}
	&{\centering {\sf 2.66}}
\\\hline
\end{longtable}
\newpage

\begin{center}
\bf{FIGURE CAPTIONS}
\end{center}
{\bf Fig. 1}. The ground state structures of neutral and charged 
SiH$_{m}$ ( m = 2-4) clusters.  Also the structures 
of SiH$_{5}$$^{+}$ and SiH$_{5}$$^{-}$ clusters are shown. All the
structures have been obtained using Gaussian98 program employing
B3LYP/6-311G**. The 
gray ball represents Si atom and small black ball represents 
H atom. Note that Si and H atoms are connected by bonds. In addition 
to position of atoms, the non-bonded electron density distribution is 
also shown schematically by unconnected half , open  and filled 
circles for neutral, positive and 
negative clusters respectively. The unconnected open and filled 
circles respectively   
represent  the decrease and increase in 
the non-bonded electron density.  
\\
{\bf Fig. 2}. Same as in Fig. 1 except for 
 neutral and charged 
Si$_{2}$H$_{n}$ ( n = 1-6 ) clusters.  
\\
{\bf Fig. 3}. Valance electron density distribution of neutral and 
charged SiH$_{2}$ clusters. The iso-contour 
value of 0.07 is chosen 
to show clearly the density distribution variation between  the neutral and 
charged SiH$_{2}$. The locations of Si and H atoms are also indicated.  
\\
{\bf Fig. 4}.  ( a ) Si-H,  ( b ) Si-H-Si and ( c ) Si-Si bond 
lengths ( $\AA$ )  
of neutral and charged  
Si$_{n}$H$_{m}$(n=1 or 2, m = 0-6 ) clusters as a function of 
cluster size. Note that for Si-H-Si bond, it is Si-H distance 
which is plotted.The circles connected by thick 
line represents the bond lengths of neutral clusters, 
the square connected by broken line for positively charged  
and triangles connected by doted line for negatively charged 
clusters.  
\\
{\bf Fig. 5}. Iso-contour distribution of HOMO in  Si$_{2}$H$_{4}$, 
Si$_{2}$H$_{3}$ and Si$_{2}$H$_{3}$$^{+}$. 
\\
{\bf Fig. 6}. The total energy of SiH$_{3}$, SiH$_{3}^{-}$ and SiH$_{3}^{+}$ 
as a function of Si distance from the center of inversion. For each
cluster the total energy is given with respect to its ground state energy. 
All these
set of calculations have been performed by varying 
Si distance along symmetry axis and freezing H atoms.  Since H 
atoms are frozen in the
present set of calculations,  for each cluster the total energy 
at the inversion center shown in the figure  is slightly  overestimated in 
comparison with the fully geometry optimized energy referred in the text.  
Note that 0.0$\AA$ means that Si atom is at the inversion center and 
the structure of the cluster is planar. 


\begin{thebibliography}{99}
\bibitem{foot1} The computational procedure for our CPMD
calculation on SiH$_{m}$ and Si$_{2}$H$_{m}$ clusters is similar to that of
our earlier CPMD calculations done on Si$_{m}$ (m=2-10) and
Si$_{m}$H ( m=1-10) clusters\cite{BalaPRB01v64}.
\bibitem{foot2}Here we would like to point out the difficulties in 
performing plane wave calculations for negatively charged 
clusters. We have performed 
plane wave pseudopotential calculations as implemented in
VASP package\cite{KressePRB96v54} for positive and negative
charged clusters. VASP set of calculations have been performed 
with ultrasoft pseudo-potential\cite{KressePRB96v54}
using  local density 
approximation(LDA)\cite{CeperleyPRL80v45,PerdewPRB81v23}.  We
have used  150 eV Ecut for neutral and
charged clusters. Calculations on neutral as well as charged
clusters have been performed using a simple cubic supercell
of 20$\AA$ length. This length is sufficiently big enough to isolate the
clusters  from their supercell image\cite{KumarPRL03v90}. 
Moreover VASP handles the charged systems by applying a background charge
to maintain the charge neutrality and by adding dipole and quadrupole 
corrections\cite{MakovPRB95v51,NeugebauerPRB92v46}.  Using the big 
supercell with the added corrections we find that the VASP calculated
total energies and structures of positive clusters are good while 
those of  negative clusters are not good. In the 
negatively charged clusters the upper most electron
is highly diffusive in nature. Therefore, negative clusters require
an even larger supercell. Since Gaussian package calculations are
performed in real space, the image problem which occurs in supercell
technique can be avoided. All the values reported for the clusters 
in this article have been obtained using Gaussian package.
\bibitem{foot3} The symmetry breaking of charged SiH$_{4}$ clusters and 
their vibrational properties have been 
discussed earlier\cite{BalaPRA03v69}. For the sake of completeness
 we have included some results in this paper.
\bibitem{foot4}The population analysis gives
charges on Si and H in SiH$_{2}^{-}$ respectively as -0.48e and
-0.26e while for neutral SiH$_{2}$ the corresponding charges
are 0.30e and -0.15e. The population
numbers given here and in the other parts of text 
refers to the amount of charge on each atom 
with respect to nuclear charge obtained  from natural population
analysis using B3LYP/6-311G** calculations.
\end{thebibliography}
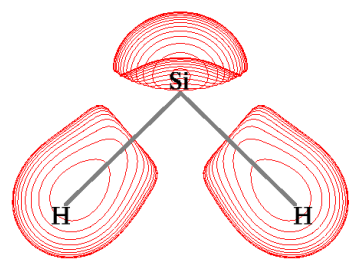
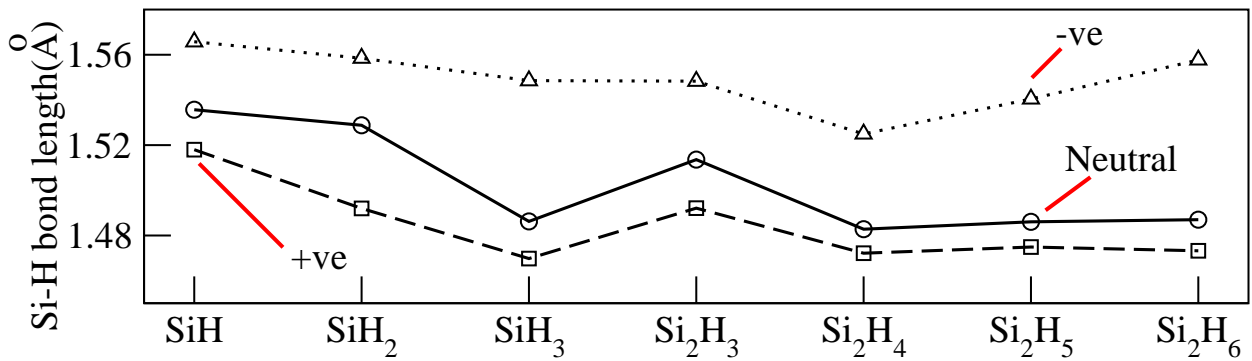
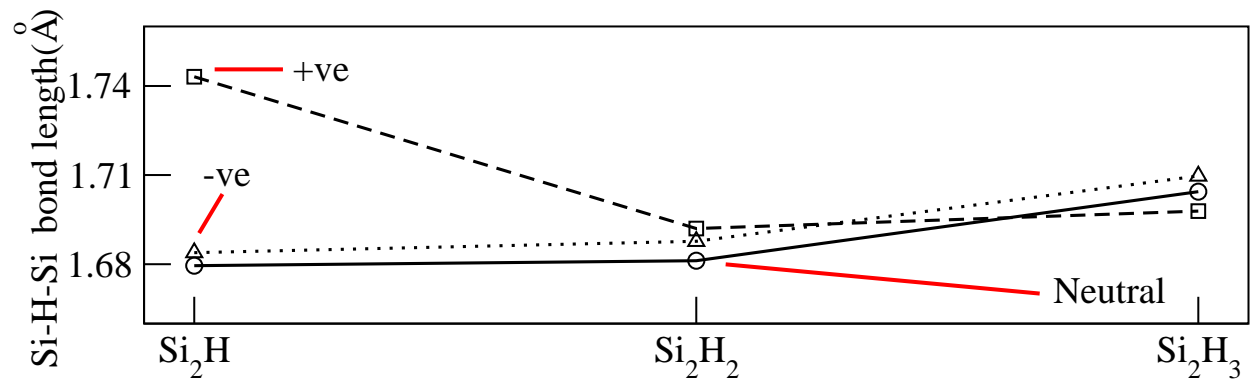
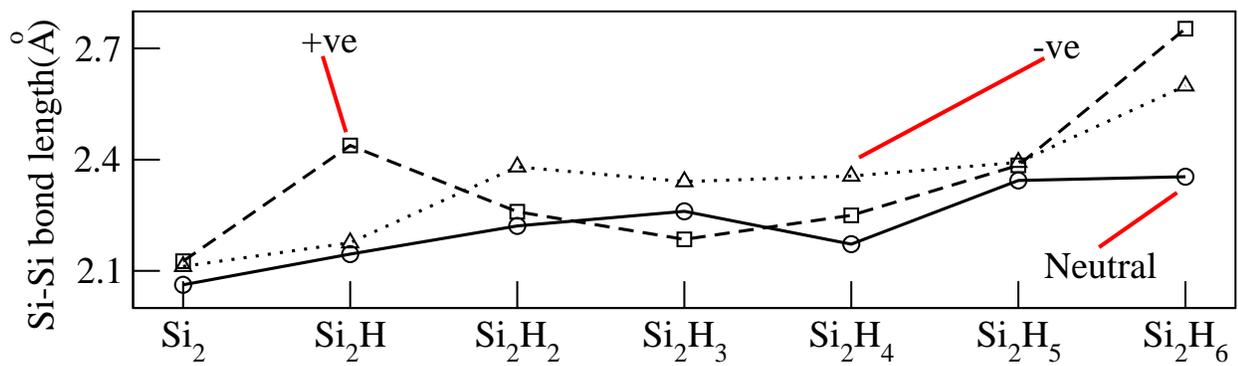
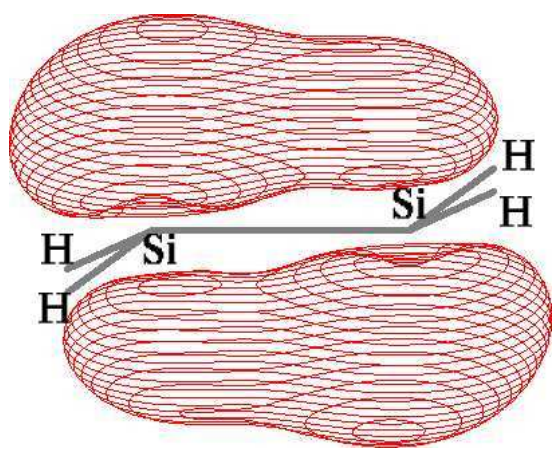
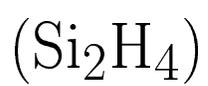
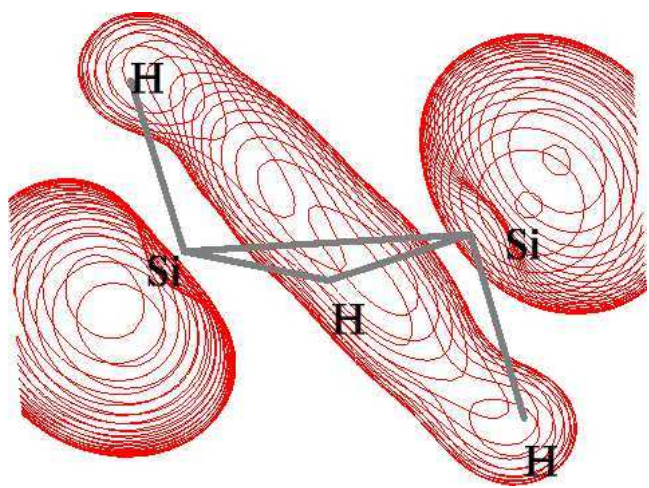
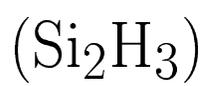
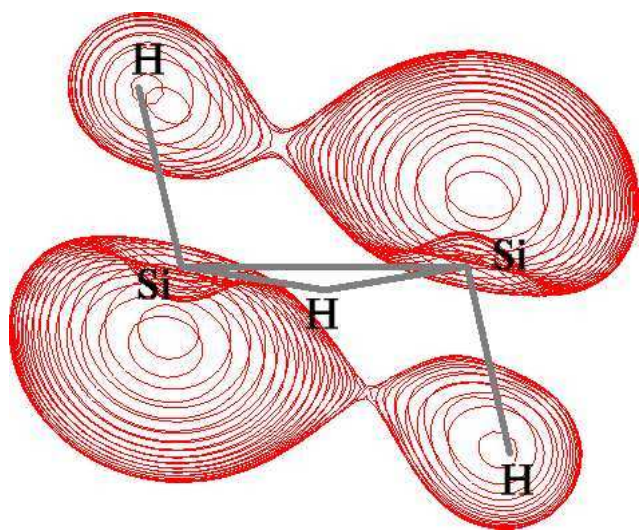
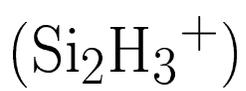
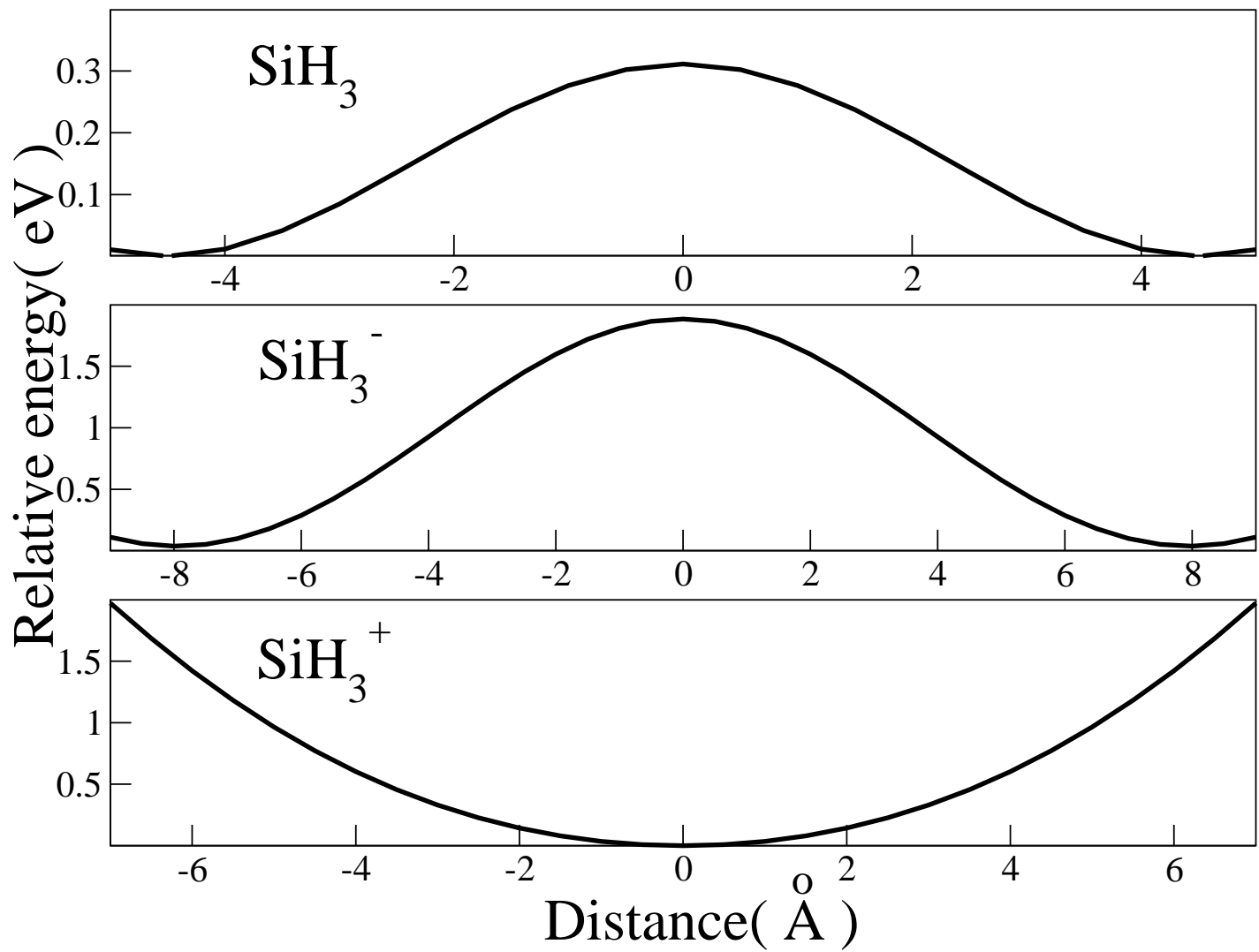
\end{document}